\documentclass{PoS}

\usepackage{graphicx}
\usepackage{amssymb}
\usepackage{amsmath}

\newcommand{\beq}{\begin{eqnarray}}
\newcommand{\eeq}{\end{eqnarray}}

\title{QCD Thermodynamics on the Lattice from the Gradient Flow}

\ShortTitle{QCD Thermodynamics on the Lattice from the Gradient Flow}

\author{\speaker{Etsuko Itou} 
\thanks{Numerical simulation for this study was carried out on Hitachi SR16000 
and IBM System Blue Gene Solution at KEK under its Large-Scale Simulation Program
(No. 14/15-11), Hitachi SR16000 at YITP Kyoto University and NEC SX-8, SX-9 and SX-ACE at RCNP Osaka University.  
S. A. is supported in part by the Grant-in-Aid of the Japanese Ministry of Education, Sciences and Technology, Sports and Culture (MEXT) for Scientific Research (No. JP16H03978),
by a priority issue (Elucidation of the fundamental laws and evolution of the universe) to be tackled by using Post ``K" Computer,
and by Joint Institute for Computational Fundamental Science (JICFuS).
This work was partially supported by RIKEN iTHES Project.}\\
        Research Center for Nuclear Physics (RCNP), Osaka University\\
        E-mail: \email{itou@yukawa.kyoto-u.ac.jp}}

\author{Sinya Aoki\\
        Yukawa Institute for Theoretical Physics, Kyoto University\\
        E-mail: \email{saoki@yukawa.kyoto-u.ac.jp}}

\abstract{To obtain the precise values of the bulk quantities and transport coefficients in quark-gluon-plasma phase, we propose that a direct calculation of the renormalized energy-momentum tensor (EMT) on the lattice using the gradient flow.
From one-point function of EMT, authors in Ref.~\cite{Asakawa:2013laa} obtained the interaction measure and thermal entropy. The results are consistent with the one obtained by the integral method.
Based on the success, we try to measure the two-point function of EMT, which is related to the transport coefficients.
Advantages of our method are (1) a clear signal because of the smearing effects of the gradient flow and (2) no need to calculate the wave function renormalization of EMT.
In addition, we give a short remark on a comparison of the numerical cost between the positive- and adjoint-flow methods for fermions, needed to obtain the EMT in the (2+1) flavor QCD.
  }

\FullConference{The 26th International Nuclear Physics Conference\\
		11-16 September, 2016\\
		Adelaide, Australia}
		
\begin{document}

\section{Introduction: energy-momentum tensor on lattice and gradient flow}
One of the most important tasks of the present heavy-ion physics is to determine the thermal properties of the quark-gluon-plasma (QGP) phase.
In this context, it is necessary to precisely calculate the thermal quantities, namely entropy, pressure and shear viscosity.
In our works, we try to determine these quantities directly from the calculations of the energy-momentum tensor (EMT).

Measurements of the EMT using the lattice numerical simulation have at least two difficulties:
One is a conceptual difficulty. The lattice regularization manifestly breaks the general covariance, while EMT is a generator of the  corresponding invariance.
The other one is a numerical cost. 
Since the signal of two-point function for the EMT operator becomes very noisy due to 
its nonzero vacuum expectation value, it is too costly to determine its renormalization factor.

In this work, we obtain the EMT using a new technique~\cite{Asakawa:2013laa} based on the small flow-time expansion  of the Yang-Mills gradient flow~\cite{Luscher:2010iy, Suzuki:2013gza}.
Advantages of the usage of the gradient flow are following:
\begin{itemize}
\item At finite flow-time, we can define the ``correctly renormalized EMT" from lattice data in the continuum limit
\item Signals become much better because of the smearing effects of the gradient flow
\item It is not necessary to calculate the wave function renormalization of the EMT operator thank to its UV finiteness (in quenched QCD)\cite{Luscher:2011bx}
\end{itemize}

In this proceeding, we briefly review of the basic idea to obtain the bulk thermal quantities, namely interaction measure (trace anomaly) and thermal entropy density, from direct calculation of the one-point function of EMT.
Next, we show the two-point function of EMT, and its flow-time dependence.
In the last section, we give a short remark for the numerical costs in the case of the(2+1) flavor QCD. We compare the numerical costs between the positive- and adjoint-flow methods proposed in Ref.~\cite{Luscher:2013cpa} to solve the gradient flow for fermions.

\section{Review: thermal quantities from one-point function of EMT}
In Ref.~\cite{Asakawa:2013laa}, one of authors (E.I.) obtained the integration measure and thermal entropy of the pure Yang-Mills theory in finite temperature from the direct calculation of EMT on the lattice.
The key relationship is given in Ref.~\cite{Suzuki:2013gza} for quenched QCD in small flow-time expansion:
the correctly-normalized EMT can be defined by
\begin{align}
   T_{\mu\nu}^R(x)
   =\lim_{t\to0}\left\{\frac{1}{\alpha_U(t)}U_{\mu\nu}(t,x)
   +\frac{\delta_{\mu\nu}}{4\alpha_E(t)}
   \left[E(t,x)-\left\langle E(t,x)\right\rangle_0 \right]\right\},
\label{eq:(4)}
\end{align}
where $\langle\cdot\rangle_0$ is vacuum expectation value (v.e.v.) and $T_{\mu\nu}^R(x)$
is the correctly-normalized conserved EMT with its v.e.v.
subtracted. 
Here $U_{\mu \nu}(t,x)$ and $E(t,x)$ denotes gauge-invariant local products of dimension~$4$ 
and they are UV finite for the positive flow-time ($t>0$).
Explicitly, 
$U_{\mu\nu}(t,x)\equiv G_{\mu\rho}(t,x)G_{\nu\rho}(t,x)
-\delta_{\mu\nu} E(t,x)$
and~$E(t,x)\equiv\frac{1}{4}G_{\mu\nu}(t,x)G_{\mu\nu}(t,x)$.
Here $G_{\mu \nu}$ represents the field strength constructed by the flowed gauge field ($B_{\mu}(t,x)$), that is a solution to the gradient flow equation as
\beq
\partial_t B_{\mu} = D_{\nu} G_{\nu \mu}, ~~~~ B_{\mu} (t=0,x) = A_{\mu}(x),
\eeq
where $A_{\mu}(x)$ denotes the original quantum gauge field variable.

The contributions from the operators of dimension~$6$ or higher are suppressed for small~$t$, and the coefficients $\alpha_U,\alpha_E$ are calculated perturbatively in \cite{Suzuki:2013gza}.

In paper~\cite{Asakawa:2013laa}, we perform the numerical simulation and obtain the bulk quantity in the quenched QCD.
We utilize the Wilson plaquette gauge action and $3$-set of ($\beta, N_\tau$) for one fixed physical-temperature to take the continuum limit. Here $\beta=6/g_0^2$ denotes the lattice bare coupling constant and has one-to-one correspondence with a lattice spacing~\cite{Guagnelli:1998ud}.
The number of gauge configurations for the measurements at each lattice parameter is only~$100$--$300$.
Statistical errors are estimated by the jackknife method.

\begin{figure}[h]
\begin{center}
\includegraphics[scale=0.4]{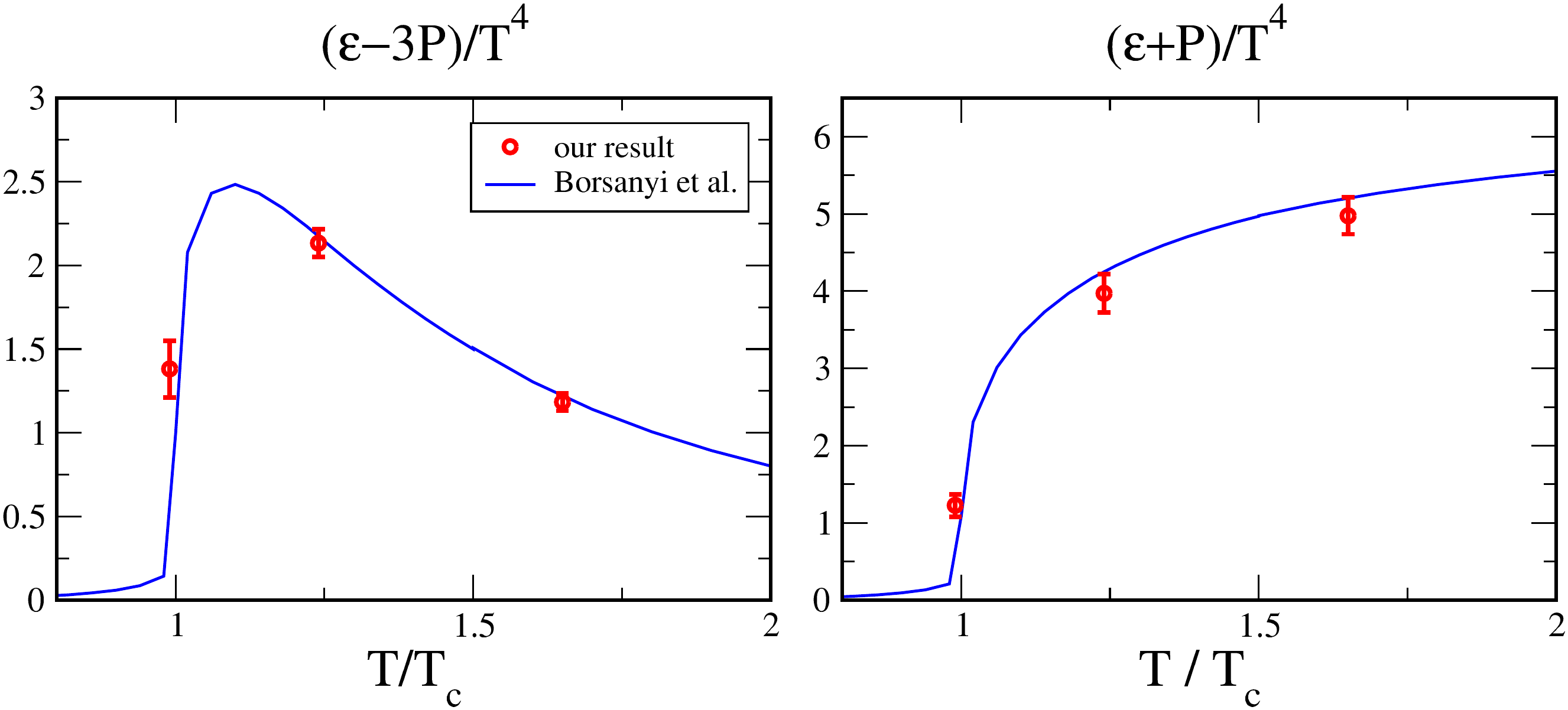}
\caption{Continuum limit of the interaction measure and entropy density
obtained by the gradient flow for~$T/T_c=1.65$, $1.24$, and~$0.99$ obtained in Ref.~\cite{Asakawa:2013laa}. Blue solid lines are results of~Ref.~\cite{Borsanyi:2012ve} obtained by the integral method.
}
\label{fig:quenched-result}
\end{center}
\end{figure}
In Fig.~\ref{fig:quenched-result}, $(\epsilon - 3P)/T^4 = \langle \sum_{\mu=1}^4  T_{\mu \mu} \rangle /T^4$ and~$(\epsilon + P)/T^4 = \langle T_{11} -T_{44} \rangle  /T^4 $ are plotted after taking the continuum limit for $T/T_c=1.65$, $1.24$, and~$0.99$, where $\epsilon, P$ denote energy density and pressure, respectively. 
For comparison, results of~Ref.~\cite{Borsanyi:2012ve} obtained by the integral method
are shown by blue solid lines in Fig.~\ref{fig:quenched-result}. The results of the two different approaches are consistent
with each other within statistical errors.

The integral method essentially calculate the free energy of thermodynamics, and is based on the macroscopic picture in finite-temperature QCD.
On the other hand, our method is based on the microscopic picture, namely 
the quantum field theory.
It is the first numerical confirmation of the consistency between micro- and macro-scopic pictures of the QGP phase in (quenched) QCD.

\section{Two-point function of EMT}
\subsection{Transport coefficients from EMT}
We now move on the calculation of the two-point function of EMT.
It is related to the shear and bulk viscosity, and here we focus on the former one, which is given by the correlation function of $T_{12}$ component.
The Euclidean correlator, which can be measured on the lattice, is defined as
\beq
C(\tau) = \frac{1}{T^5} \int d \vec{x} \langle T_{12} (0, \vec{0}) T_{12}(\tau, \vec{x}) \rangle,
\eeq 
which is expressed in terms of the corresponding  spectral functions ($\rho(\omega)$) as
\beq
C(\tau) = \frac{1}{T^5} \int_{0}^{\infty} d \omega \rho(\omega) \frac{\cosh \omega \left( \frac{1}{2T} - \tau \right)}{\sinh(\frac{\omega}{2T})}.
\eeq
The shear viscosity in QGP phase is then given by
\beq
\eta(T) = \pi \left. \frac{d \rho}{d\omega} \right|_{\omega=0}.
\eeq

There are several works~\cite{Nakamura:2004sy, Meyer:2007ic, Mages:2015rea}, where the correlation function of EMT are calculated on lattice.
In these works, we explained before, there exit at least two difficulties, the renormalization of the lattice bare EMT operator and the bad signal to noise ratio of the quantity.
In Ref.~\cite{Meyer:2007ic}, the author introduced the one-loop lattice-perturbative $Z(g_0)$ factor to define the renormalized EMT, while in Ref.~\cite{Mages:2015rea} they estimate $Z$-factor for a diagonal component of EMT from the thermal entropy shown in Fig.~\ref{fig:quenched-result}.
Our method do not need the calculation of $Z$-factor if we use the renormalized coupling constant in the coefficient $\alpha_U, \alpha_E$, thank to the UV finiteness of the flowed-composite operators.
The signal to noise ratio is also drastically improved by the gradient flow because of its smearing effects, as we will show.

\subsection{Lattice setup}
We consider the Wilson plaquette gauge action on $N_\tau = 8, 10, 12$ and $16$ lattices with a fixed $N_s/N_\tau =4$.
The lattice bare coupling constant $\beta$ for each $N_\tau$ is tuned to realize the temperature $T=1.65T_c$.
The corresponding $\beta$ for each $N_\tau$ is determined by the relation in ALPHA Collaboration~\cite{Guagnelli:1998ud}.

Gauge configurations are generated without dynamical fermions by the pseudo-heatbath algorithm with the over-relaxation.
We call one pseudo-heatbath
update sweep plus several over-relaxation sweeps
as a Sweep. To eliminate the autocorrelation, we take
$200$ Sweeps between measurements. The number of
gauge configurations for the measurements at finite T
is $3000$, $5000$, $2500$ and $800$ for $N_\tau=8,10,12$ and $16$, respectively. 
Statistical errors are estimated by the jackknife method.

\subsection{Results}
Firstly, we show the improvement of the statistical uncertainty by the usage of the gradient flow.
For the correlation function of $(T_{11}-T_{22})/2$ operator, the similar results are reported in Ref.~\cite{Mages:2015rea}.
Here, we show the results for $T_{12}$ correlator, which is equal to the one for $(T_{11}-T_{22})/2$ in the continuum limit.

\begin{figure}[h]
\begin{center}
\includegraphics[scale=0.5]{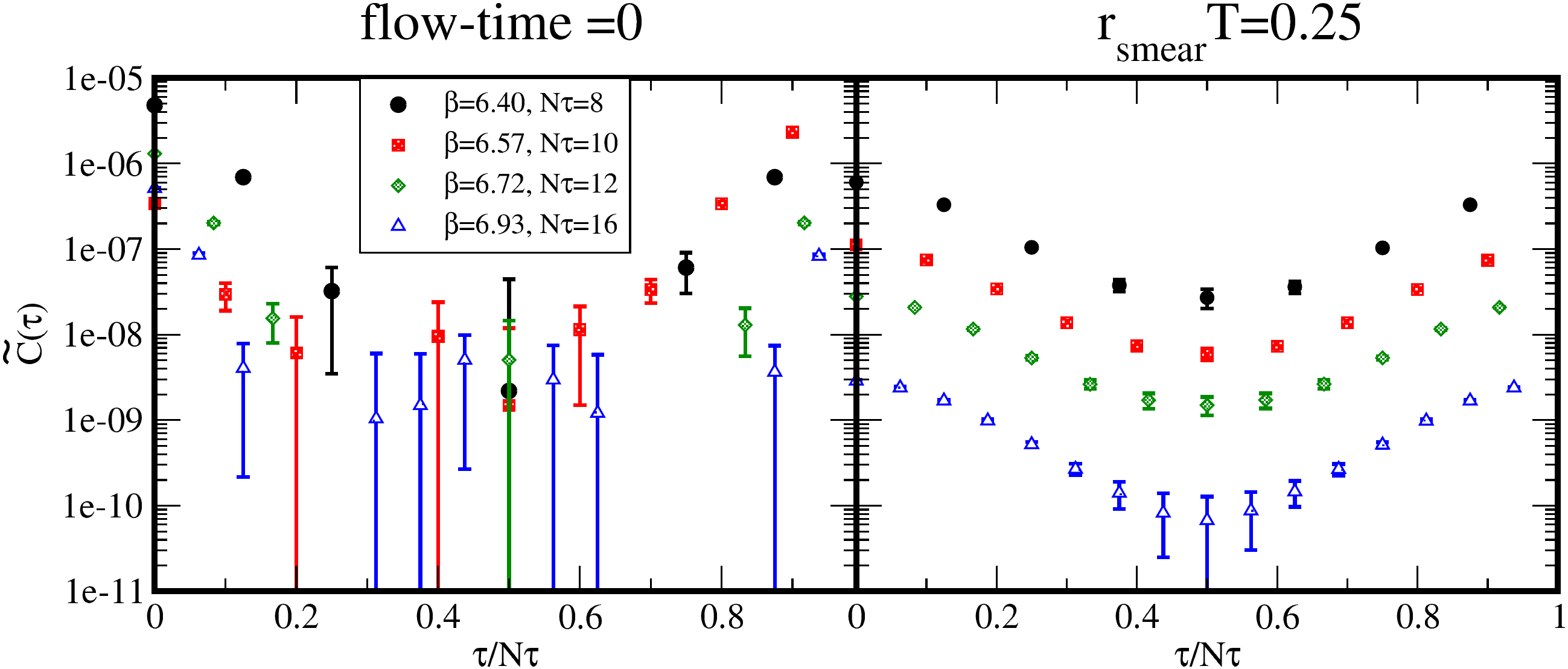}
\caption{Correlation function of $U_{12}$ operator, $\tilde{C}(\tau)= \langle \int d \vec{x} \langle U_{12} (0, \vec{0}) U_{12}(\tau, \vec{x}) \rangle$, without the gradient flow (left panel) and with the gradient flow (right panel), where the number of the measured configuration for each color in both panels is the same.}
\label{fig:plot-U12U12}
\end{center}
\end{figure}

In Fig.~\ref{fig:plot-U12U12}, we plot the correlation function of $U_{12}$ operator, which is defined by the clover leaf on the lattice, without the gradient flow (right panel) and with the gradient flow (left panel). Here the number of the measured configurations for each color in both panels is the same.
Although the data should be positive by definition and indeed so in Ref.~\cite{Nakamura:2004sy} with high statistics, some un-flowed data in Fig.~\ref{fig:plot-U12U12} become negative due to large noises in this statistics.
On the other hand, at the finite flow-time (we take $r_{\mathrm{smear}}T=0.25$ with $r_{\mathrm{smear}}=\sqrt{8t}$), the correlation function is positive at all $\tau$ despite low statistics, demonstrating that signals are highly improved.

\begin{figure}[h]
\begin{center}
\includegraphics[scale=0.45]{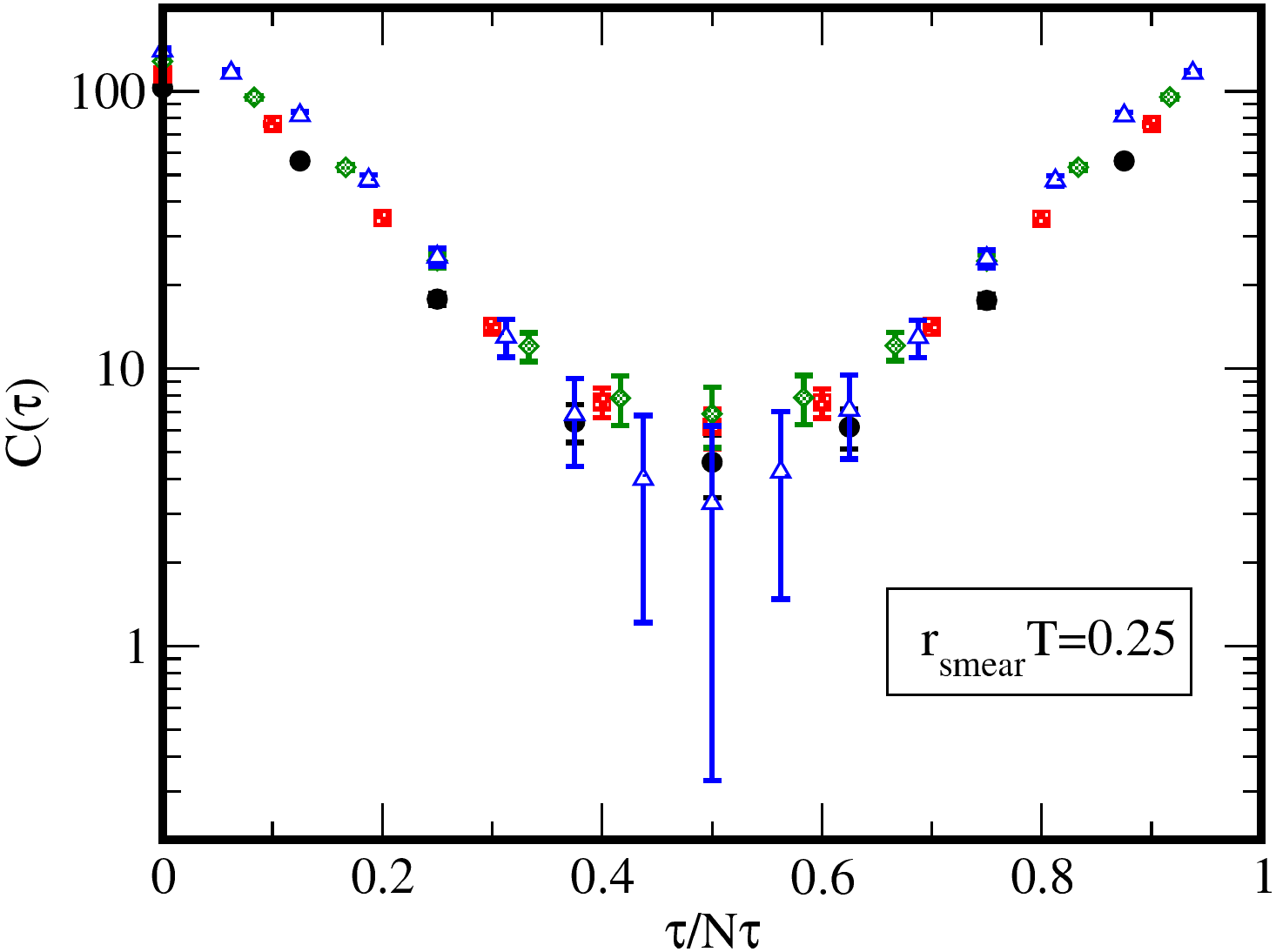}
\caption{Correlation function of the renormalized $T_{12}$ operator at $r_{\mathrm{smear}}T=0.25$ for $N_\tau=8,10,12$ and $16$ lattices.
}
\label{fig:T12T12-disc-error}
\end{center}
\end{figure}

Figure~\ref{fig:T12T12-disc-error} shows the correlation function for the renormalized $T_{12}$ operator at finite flow-time, which includes the $\beta$ dependent coefficient $1/\alpha_U^2$.
The discrepancy among data for different $N_\tau$ comes from the discretization error of the $T_{12}$ correlator in our formulation.
We found that it is small and looks under-controlled in whole range of the imaginary-time.

We also show the flow-time dependence of the shape of the correlation function.
\begin{figure}[h]
\begin{center}
\includegraphics[scale=0.45]{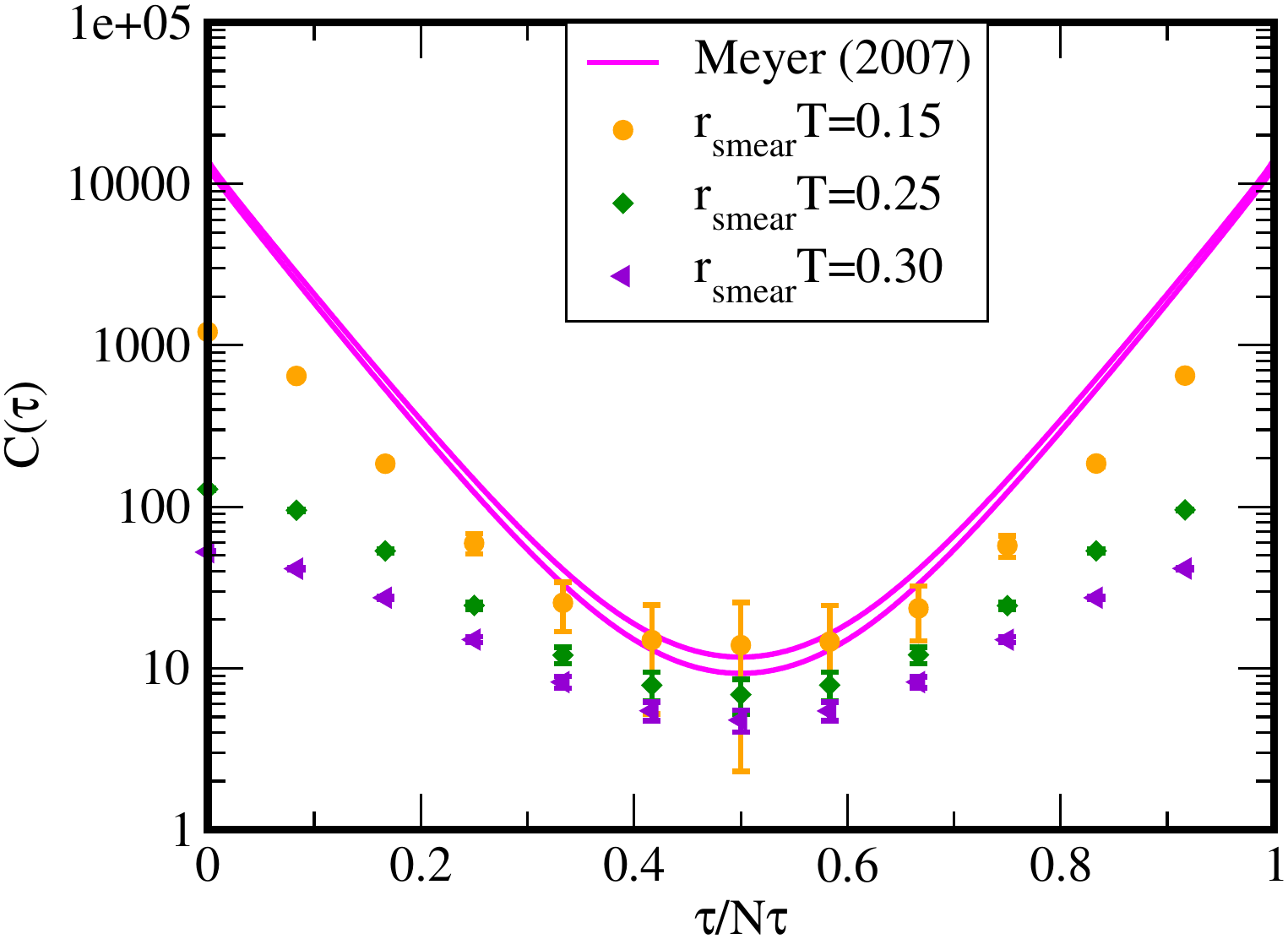}
\caption{Flow-time dependence of the $T_{12}$ correlator. Here the data of $\beta=6.72, N_\tau=12$ are plotted. Magenta curves denotes the best-fit in Ref.~\cite{Meyer:2007ic} with 1-$\sigma$ error of the fitting parameters in the Breit-Wigner fit ansatz .
}
\label{fig:T12T12-corr-flowtime-deps}
\end{center}
\end{figure}
In Fig.~\ref{fig:T12T12-corr-flowtime-deps}, we found that the slope of the correlation function becomes milder in the longer flow-time.
This is natural, since in the large flow-time limit the operator of $T_{12}$ are smeared in whole temporal direction and then the correlation becomes a constant as a function of $\tau$.
We also plot the best-fit function of the un-flowed data in Ref.~\cite{Meyer:2007ic} by obtaining the Breit-Wigner fit ansatz.
Although the data in Ref.~\cite{Meyer:2007ic} is $(T_{11}-T_{22})/2$ correlator with the different renormalization process, the magenta curve might be consistent with the $t \rightarrow 0$ limit of our flowed-data.

\subsection{Outlook and future plan}
Finally, we put our future plans to obtain the shear viscosity using our method.
One possible method  is to extract the spectral function $\rho(\omega)$  from the correlation function by the fit with some assumptions on the functional form of $\rho(\omega)$.
In this case,
we have to remove data at $\tau/N_\tau \le r_{\mathrm{smear}}T$, which suffer from the (over-)smearing effects.
From $\rho(\omega)$, we obtain the $\eta/T^3$ at several finite flow-times.
Since the flow-time dependence of the EMT two-point function looks stronger than that of the one-point function (See Fig.~1 in Ref.~\cite{Asakawa:2013laa}), we expect that the flow-time dependence of $\eta/T^3$ is also larger, so that a careful estimation is necessary.

\section{(2+1)-flavor QCD calculation in the positive- and adjoint-flow}
\subsection{Positive-flow and adjoint-flow to solve the gradient flow equation for fermions }
In this section, we would like to give a short remark on the calculation of the EMT components including the fermion fields for full QCD system.
The basic strategy and some preliminary results before taking the continuum limit are provided in Ref.~\cite{Itou:2015gxx, Taniguchi:2016ofw}.
In the numerical calculation in these papers, the simulation cost to solve the gradient flow for fermions using the adjoint-flow method are high. 
In this proceedings, we compere the numerical costs between two methods, namely the adjoint-flow and positive-flow\cite{Luscher:2013cpa}, when we solve the gradient flow equation for the fermion to obtain the thermal quantities.

The small flow-time formula for the definition of the EMT in full QCD system is given in Ref.~\cite{Makino:2014taa}.
To obtain the EMT, we have to calculate following two-types of expectation values.
\beq
t^f_{\mu \nu}(t) &\equiv& \frac{1}{N_\Gamma} \sum_x \langle \bar{\chi}^f (t,x) \gamma_\mu \left( D_\nu - \overset{\leftarrow}{D}_\nu  \right) \chi^f (t,x) \rangle,\label{eq:t-tensor} \\
s^f(t) &\equiv& \frac{1}{N_\Gamma} \sum_x \langle \bar{\chi}^f (t,x) \chi^f (t,x) \rangle,\label{eq:s-tensor}
\eeq
where $N_\Gamma$ is the lattice volume in lattice unit, $f$ denotes a label of the quark flavor and $\Delta \chi^f (t,a)=D_\mu D_\mu \chi^f(t,x)$.
$\chi(t,x),\bar{\chi}(t,x)$ are solutions of the fermion flow equation, which is given by
\beq
\partial_t \chi^f(t,x) &=& \Delta \chi^f(t,x), ~~~\chi^f(t=0,x)=\psi^f(x),\nonumber\\
\partial_t \bar{\chi}^f(t,x) &=& \bar{\chi}^f (t,x) \overset{\leftarrow}{\Delta}, ~~~\bar{\chi}^f (t=0,x)=\bar{\psi}^f(t,x).
\eeq
Note that the covariant derivative refers to the flowed gauge field at the flow time $t$.

To solve this equation and obtain the expectation value of composite operators, two-types of methods are proposed in Ref.~\cite{Luscher:2013cpa}.
The first one is the  ``positive-flow" method, in which we introduce the random source field at $t=0$, while the second one is the ``adjoint-flow" method where we introduce the random source at the flow-time at which we want to measure the observables and inversely solve the equation from the finite flow-time to $t=0$.

The technical procedures of the adjoint flow and positive flow are summarized as follow.
If we consider to obtain the vacuum expectation values of EMT at $t/a^2=2.0$, as calculated in Ref.~\cite{Itou:2015gxx} using the adjoint-flow methods, we carry out the following steps.

{\bf Ad-1} Solve the gauge flow and store the flowed-link variable with small flow-time interval.

{\bf Ad-2} Generate the noise vector at the finite flow-time where we want to obtain the expectation value of the observables, and solve the backward flow from the finite flow-time to zero.

{\bf Ad-3} Calculate the propagator at $t=0$ using obtained pseudo-fermion vector in {\bf Ad-2} and calculate the expectation value of $t_{\mu \nu}(t), s (t)$.

Here, to reduce the simulation cost of {\bf Ad-2} procedure, we firstly generate and store the flowed configurations with the small interval in {\bf Ad-1} ({\it e.g.} $\Delta t/a^2=0.1$ in Ref.~\cite{Itou:2015gxx} ).
Steps {\bf Ad-1} and {\bf Ad-2} are flavor independent, so that the flavor dependent part is the calculation of the propagator in {\bf Ad-3}.
On the other hand, the actual procedure in the positive-flow method is as follows.

{\bf Po-1} Generate the noise vector  and calculate the propagator for each flavor at $t=0$.

{\bf Po-2}  Solve the flow equation for link- and fermion-fields simultaneously toward the positive flow-time direction and calculate the expectation value at finite flow-time.

\subsection{Comparison of numerical costs between the positive- and adjoint-flow}

Now, we show the simulation costs for each procedure.
All calculations have been done using $32$-MPI processes of $1$ CPU on HITACHI SR16000.
We utilize a configuration generated by the Iwasaki-gauge action and the $N_f=2+1$ $O(a)$-improved Wilson fermion. The lattice parameters of the configuration are $\beta=1.900$, $c_{SW}=1.715$, $\kappa_{u,d}=0.1368$ and $\kappa_s=0.1364$ on $32^3 \times 8$ lattice.

Table~\ref{table:simulation-cost-each} shows the computation cost for each procedure.
Here the length of flow time is fixed as $\Delta t/a^2=0.1$ for {\bf Ad-1},{\bf Ad-2} and {\bf Po-2}. 
To solve the gradient flow, we use the third-order Runge-Kutta algorithm with $\epsilon=0.01$.

In {\bf Ad-3} and {\bf Po-1}, the calculation of propagator is included, so that the computational cost of this part depends on the configuration. Here we compare them using the same configuration and the same convergence precision to solve the inverse of the Dirac operator.
\begin{table}[h]
\begin{center}
\begin{tabular}{|c|c||c|c|}
\hline
Procedure  & comp. time [sec] & Procedure  & comp. time [sec]  \\  
\hline
{\bf Ad-1}      &  14                &  {\bf Po-1}    & 323 \\
{\bf Ad-2}      &  846             &  {\bf Po-2}   &  1,022       \\
 {\bf Ad-3}     &  456             &                   &        \\
\hline
\end{tabular}
\caption{ Simulation cost for each procedure. The length of flow-time is $\Delta t/a^2=0.1$ in {\bf Ad-1},{\bf Ad-2} and {\bf Po-2}.} \label{table:simulation-cost-each}
\end{center}
\end{table}

If we calculate the expectation value at flow-time $t/a^2=2.0$, the computational cost in the positive flow for the $(2+1)$-flavor QCD is
\beq
\mbox {positive-flow :  } 21,104 \mathrm{ [sec]}.
\eeq
Here, in {\bf Po-2}, we can continuously solve the gradient flow equation from $t=0$ to $t/a^2=2.0$, so that the overhead of I/O of configurations can be reduced rather than the $20$ times that of the time shown in Table~\ref{table:simulation-cost-each}.
On the other hand, in the adjoint-flow we have to recursively carry out the procedure {\bf Ad-2} due to the backward flow is unstable.
To reduce the simulation costs and to know the flow-time dependence, for instance, we take $20$ data point between $0\le t/a^2 \le 2.0$ with the interval $\Delta t/a^2=0.1$ as shown in Ref.~\cite{Itou:2015gxx}.
We repeatedly carry out the {\bf Ad-2} procedure $210$ times, and then 
the total computational cost is
\beq
\mbox {adjoint-flow :  } 196,180 \mathrm{ [sec]}.
\eeq

If we need the longer flow-time simulation, the cost of the positive-flow is further cheaper than the one for the adjoint-flow method.


\begin{thebibliography}{99}


\bibitem{Asakawa:2013laa} 
  M.~Asakawa {\it et al.} [FlowQCD Collaboration],
  Phys.\ Rev.\ D {\bf 90}, no. 1, 011501 (2014)
  [Phys.\ Rev.\ D {\bf 92}, no. 5, 059902 (2015)]
  [arXiv:1312.7492 [hep-lat]].

\bibitem{Luscher:2010iy} 
  M.~L\"uscher,
  JHEP {\bf 1008}, 071 (2010)
  [arXiv:1006.4518 [hep-lat]].

\bibitem{Suzuki:2013gza} 
  H.~Suzuki,
  PTEP {\bf 2013}, 083B03 (2013)
  [PTEP {\bf 2015}, 079201 (2015)]
  [arXiv:1304.0533 [hep-lat]].

\bibitem{Luscher:2011bx} 
  M.~L\"uscher and P.~Weisz,
  JHEP {\bf 1102}, 051 (2011)
  [arXiv:1101.0963 [hep-th]].
  
\bibitem{Luscher:2013cpa} 
  M.~Luscher,
  JHEP {\bf 1304}, 123 (2013)
  [arXiv:1302.5246 [hep-lat]].


\bibitem{Guagnelli:1998ud} 
  M.~Guagnelli {\it et al.}  [ALPHA Collaboration],
  Nucl.\ Phys.\ B {\bf 535}, 389 (1998)
  [hep-lat/9806005].


\bibitem{Borsanyi:2012ve} 
  S.~Borsanyi, G.~Endrodi, Z.~Fodor, S.~D.~Katz and K.~K.~Szabo,
  JHEP {\bf 1207}, 056 (2012)
  [arXiv:1204.6184 [hep-lat]].
  
\bibitem{Nakamura:2004sy} 
  A.~Nakamura and S.~Sakai,
  Phys.\ Rev.\ Lett.\  {\bf 94}, 072305 (2005)
  [hep-lat/0406009].

\bibitem{Meyer:2007ic} 
  H.~B.~Meyer,
  Phys.\ Rev.\ D {\bf 76}, 101701 (2007)
  [arXiv:0704.1801 [hep-lat]].

\bibitem{Mages:2015rea} 
  S.~W.~Mages, S.~Bors\'{a}nyi, Z.~Fodor, A.~Sch\"{a}fer and K.~Szab\'{o},
  PoS LATTICE {\bf 2014}, 232 (2015).


\bibitem{Itou:2015gxx} 
  E.~Itou, H.~Suzuki, Y.~Taniguchi and T.~Umeda,
  PoS LATTICE {\bf 2015}, 303 (2016)
  [arXiv:1511.03009 [hep-lat]].

\bibitem{Taniguchi:2016ofw} 
  Y.~Taniguchi, S.~Ejiri, R.~Iwami, K.~Kanaya, M.~Kitazawa, H.~Suzuki, T.~Umeda and N.~Wakabayashi,
  arXiv:1609.01417 [hep-lat].

\bibitem{Makino:2014taa}
H.~Makino and H.~Suzuki,
  PTEP {\bf 2014}, 063B02 (2014)
  [PTEP {\bf 2015}, 079202 (2015)]
  [arXiv:1403.4772 [hep-lat]].
  
\end{thebibliography}
\end{document}